\documentclass[11pt,fleqn]{article}
\usepackage{epsfig}
\def\greaterthansquiggle{\raise.3ex\hbox{$>$\kern-.75em\lower1ex\hbox{$\sim$}}}
\def\lessthansquiggle{\raise.3ex\hbox{$<$\kern-.75em\lower1ex\hbox{$\sim$}}}

\newcommand{\la}{\label}
\newcommand{\re}{\ref}
\newcommand{\ci}{\cite}
\newcommand{\beqn}{\begin{eqnarray}}
\newcommand{\eeqn}{\end{eqnarray}}
\newcommand{\bequ}{\begin{equation}}
\newcommand{\eequ}{\end{equation}}
\newcommand{\bsl}{\begin{sloppypar}}
\newcommand{\esl}{\end{sloppypar}}

\begin{document}
\setlength{\unitlength}{1cm}
\null
\hfill DESY 00-100\\
\null
\hfill WUE-ITP-2000-021\\
\null
\hfill hep-ph/0012229\\
\vskip .4cm
\begin{center}
{\Large \bf Implications of CP and CPT for 
\\[.4em]
production and decay of Majorana fermions}\\[.4em]
\vskip 1.5em

{\large
{\sc 
G.~Moortgat--Pick$^{a}$\footnote{e-mail:
    gudrid@mail.desy.de},
H. Fraas$^{b}$\footnote{e-mail:
    fraas@physik.uni-wuerzburg.de}
}}\\[3ex]                     
\end{center}
{\footnotesize \it
$^{b}$ DESY, Deutsches Elektronen--Synchrotron, D--22603 Hamburg, Germany}\\
{\footnotesize \it
$^{a}$ Institut f\"ur Theoretische Physik, Universit\"at
W\"urzburg, D--97074 W\"urzburg, Germany}\\
\vskip .5em
\par
\vskip .4cm

\begin{abstract}
The consequences of CP-- and CPT--invariance for  
production and subsequent decay of Majorana    
fermions are analytically studied. 
We derive general
symmetry relations for the spin density matrix for 
production of 
Majorana fermions by polarized fermion--antifermion annihilation which  
allow to distinguish Majorana from Dirac fermions.
We discuss the influence of the spin correlations on
energy and angular distributions of the decay products.
Numerical results are shown for the production of neutralinos and charginos 
with subsequent leptonic decay
at a future linear collider with longitudinally polarized beams. 
\end{abstract}
\vspace{1em}
\hfill

\newpage
\section{Introduction}
\label{sec:1}
For production and subsequent decay of short living particles
quantum mechanical interference effects between the various helicity
amplitudes in general preclude the factorization of the 
differential cross section into production and decay. 

In \ci{Tata} it has been
demonstrated that for suitably chosen Lorentz invariant variables
certain distributions of the decay products 
are independent of spin correlations between production and decay.
In this paper we derive that in the case of Majorana
fermions additional distributions 
are independent of spin correlations due to CPT-- and CP--invariance. 
Moreover in this study general beam polarization is included.

The production cross section of Majorana fermions is forward--backward
(FB) symmetric if CP is conserved. 
In leading order pertubation theory if there are no absorptive effects
this also follows from CPT invariance of the S--matrix \cite{Petkov}.
From CPT and CP invariance we derive 
symmetry properties of the  
production spin density matrix of Majorana fermions with polarized beams.  
These symmetry properties have consequences for the factorization 
of angular and energy distributions in the lab frame, which opens a
possibility
to distinguish between Majorana and Dirac fermions.

In the following section 2 we give details of the spin density matrix 
formalism. In section 3 we discuss the consequences of
CPT and CP invariance for the 
production density matrix of two different Majorana fermions. 
In section 4 we study the 
consequences for angular distributions and the energy spectra of the 
decay leptons. Numerical results are given in section~5 for 
differential cross sections of production and leptonic decay of neutralinos and
charginos in $e^+ e^-$ annihilation with longitudinally polarized beams in the 
framework of the Constrained Minimal Supersymmetric Standard Model (CMSSM). 
\section{Spin correlations between production and decay}\la{sec:2}
We consider pair production of Majorana or Dirac fermions in
fermion--antifermion annihilation and the subsequent three--particle decay.
The helicity amplitudes for the production processes 
\bequ
f(p_1,\lambda_1) \bar{f}(p_2, \lambda_2)\to f_i(p_i, \lambda_i) 
f_j(p_j, \lambda_j)
\la{eq2_1}
\eequ
are denoted by $T_{P, \lambda_1 \lambda_2}^{\lambda_i\lambda_j}$ 
and those for the decay processes
\beqn
&&
f_{i}(p_i, \lambda_i)\to f_{i1}(p_{i1})f_{i2}(p_{i2})f_{i3}(p_{i3}),
\la{eq2_2a}\\
&&
f_{j}(p_j, \lambda_j)\to f_{j1}(p_{j1})f_{j2}(p_{j2})f_{j3}(p_{i3})
\la{eq2_2b}
\eeqn
by $T_{D,\lambda_i}$ and $T_{D,\lambda_j}$. In the notation
for all produced particles, (\re{eq2_1})--(\re{eq2_2b}), 
it is not distinguished between fermions and antifermions.
The helicities of the decay products are suppressed.

For polarized beams we use the spin density matrices
$\rho(f)=\frac{1}{2}(1+P^i_f\sigma^i)$ and
$\rho(\bar{f})=\frac{1}{2}(1+P^i_{\bar{f}}\sigma^i)$, where 
$P^1_{f}$, $P^2_f$, $P^3_f\equiv P_f$ 
($P^1_{\bar{f}}$, $P^2_{\bar{f}}$, $P^3_{\bar{f}}\equiv P_{\bar{f}}$)
is the transverse polarization of $f$ ($\bar{f}$) in the production
plane, the polarization normal to the production plane and the longitudinal
polarization. 

The amplitude squared of the combined process of production and decay is:
\mathindent0cm
\bequ
|T|^2=|\Delta(f_i)|^2 |\Delta(f_j)|^2 
\rho_P^{\lambda_i \lambda_j, \lambda^{'}_i \lambda^{'}_j} 
\rho_{D, \lambda^{'}_i\lambda_i}\rho_{D,\lambda_j^{'} \lambda_j},
\quad\mbox{summed over helicities}.
\la{eq4_4e}
\eequ
It is composed from the (unnormalized) spin density production matrix
$\rho_P^{\lambda_i\lambda_j, \lambda_i^{'}\lambda_j^{'}}=
\rho(f)_{\lambda'_1\lambda_1}\rho(\bar{f})_{\lambda'_2 \lambda_2}
T_{P, \lambda_1 \lambda_2}^{\lambda_i\lambda_j}
T_{P, \lambda'_1 \lambda'_2}^{\lambda_i^{'}\lambda_j^{'}*}$,
the decay matrices
$\rho_{D, \lambda_i^{'} \lambda_i}=T_{D, \lambda_i}T_{D,\lambda_i^{'}}^{*}$
and 
 $\rho_{D, \lambda_j^{'} \lambda_j}=T_{D, \lambda_j}T_{D,\lambda_j^{'}}^{*}$
and the propagators 
$\Delta(f_{k})=1/[p^2_{k}-m_{k}^2+i m_{k} \Gamma_{k}]$, $k=i,j$.
Here $p_{k}^2$, $m_{k}$ and $\Gamma_{k}$ denote the 
four--momentum squared, mass and total width of the fermion 
$f_{k}$. For these propagators we use the narrow--width 
approximation. 

We introduce for $f_i$ ($f_j$)
three spacelike polarization vectors
 $s^{a\mu}(f_i)$
($s^{b\mu}(f_j)$) which together with $p_i^{\mu}/m_i$
($p_j^{\mu}/m_j$) form an orthonormal set \ci{Haber}.
Then the spin density matrix of production and the decay matrices can
be expanded in terms of Pauli matrices $\sigma^a$: 
\beqn 
\rho_P^{\lambda_i\lambda_j, \lambda_i^{'}\lambda_j^{'}}&=&
\delta_{\lambda_i\lambda_i^{'}} \delta_{\lambda_j\lambda_j^{'}}
P(f_i f_j)
+\delta_{\lambda_j\lambda_j^{'}}\sum_{a=1}^3
\sigma^a_{\lambda_i\lambda_i^{'}}\Sigma^a_P(f_i)
\nonumber\\ &&
+\delta_{\lambda_i\lambda_i^{'}}\sum_{b=1}^3
\sigma^b_{\lambda_j\lambda_j^{'}}\Sigma^b_P(f_j)
+\sum_{a,b=1}^3\sigma^a_{\lambda_i\lambda_i^{'}}
\sigma^b_{\lambda_j\lambda_j^{'}}
\Sigma^{ab}_P(f_i f_j),\la{eq4_4h}\\
\rho_{D,\lambda_i^{'}\lambda_i}&=&\delta_{\lambda_i^{'}\lambda_i}
D(f_i)+\sum_{a=1}^3
\sigma^a_{\lambda_i^{'}\lambda_i} \Sigma^a_D(f_i),\la{eq4_4i}\\
\rho_{D,\lambda_j^{'}\lambda_j}&=&\delta_{\lambda_j^{'}\lambda_j}
D(f_j)+\sum_{b=1}^3
\sigma^b_{\lambda_j^{'}\lambda_j} \Sigma^b_D(f_j).\la{eq4_4j}
\eeqn
Here $a,b=1,2,3$ refers to the 
polarization vectors of $f_i$ ($f_j$). 
In (\re{eq4_4h}) the dependence of $\rho_P$
on beam polarization has been suppressed.
We choose the 
polarization vectors such that in the lab system 
$\vec{s^{3}}(f_i)$ ($\vec{s^{3}}(f_j)$) is in the direction of 
momentum $\vec{p_i}$ ($\vec{p_j}$) and 
$\vec{s^2}(f_i)=\frac{\vec{p}_1\times \vec{p}_i}
{|\vec{p}_1\times \vec{p}_i|}=\vec{s^2}(f_j)$ 
is perpendicular to the production plane 
and $\vec{s^1}(f_i)$ ($\vec{s^1}(f_j)$) is in the production plane 
orthogonal to the momentum $\vec{p}_i$ ($\vec{p}_j$) such that
$\vec{s^1}$, $\vec{s^2}$, $\vec{s^3}$ form an orthogonal right--handed 
system.

Then 
$\Sigma^3_P(f_{i,j})/P(f_i f_j)$ 
is the longitudinal polarization,
$\Sigma^1_P(f_{i,j})/P(f_i f_j)$ is 
the transverse polarization in the scattering plane and
$\Sigma^2_P(f_{i,j})/P(f_i f_j)$ 
is the polarization perpendicular to the scattering plane.
In the following we refer to the polarization of $f_i$ ($f_j$) by
$\Sigma^a_P(f_i)$ ($\Sigma^b_P(f_j)$).
The terms $\Sigma^{ab}_P(f_i f_j)$ are due to correlations between 
the polarizations of both produced particles. 
For neutralinos and charginos
the complete analytical expressions for 
the production density matrix and for the decay matrices 
are given in \ci{Gudi_neut,Gudi_char}.

The amplitude squared $|T|^2$  
of the combined process of production and decay (\ref{eq4_4e}) can
be written as:
\beqn
|T|^2&=&4|\Delta(f_i)|^2|
\Delta(f_j)|^2
         \Big(P(f_i f_j) 
D(f_i) D(f_j)
\nonumber\\ &&
    +\sum^3_{a=1}\Sigma_P^a(f_i) 
\Sigma_D^a(f_i) 
D(f_j)
+\sum^3_{b=1}\Sigma_P^b(f_j) \Sigma_D^b(f_j)
D(f_i)\nonumber\\ &&
    +\sum^3_{a,b=1}\Sigma_P^{ab}(f_i f_j)
 \Sigma^a_D(f_i) \Sigma^b_D(f_j)\Big),
\la{eq4_5}
\eeqn
and the differential cross section is given by
\begin{equation}
d\sigma=\frac{1}{2 s}|T|^2 (2\pi)^4
\delta^4(p_1+p_2-\sum_{i} p_i) d{\rm lips}\label{eq_13},
\end{equation}
where $d{\rm lips}$ 
is the Lorentz invariant phase space element.
\section{ Constraints on the production of Majorana\\ \hspace*{.1cm} 
and Dirac fermions from
CPT and CP}\la{sec:3}
In this section we derive for polarized beams constraints 
from CPT and CP invariance for the joint
density matrix (\re{eq4_4h}) for production 
of two different Majorana fermions and of two Dirac fermions, respectively.

\bsl
All contributions from the exchange of particles $\alpha$, $\beta$
to the terms 
${\cal M}=\{P, \Sigma^a_P, \Sigma^b_P, \Sigma^{ab}_P\}$
of the spin--density matrix
$\rho_P^{\lambda_i \lambda_j, \lambda'_i \lambda'_j}$ (\ref{eq4_4h}), 
are composed
of products ${\cal C}$ of couplings, the propagators
$\Delta(\alpha)$, $\Delta(\beta)$ and 
complex functions ${\cal S}$ of momenta and
polarization vectors:
$
{\cal M}\sim Re\{{\cal C} \times \Delta(\alpha) \Delta(\beta)^{*} 
\times {\cal S} \}.
$
\esl
In the general case of complex couplings and finite widths of the 
exchanged particles the general structure can be written:
\beqn
{\cal M}
&\sim&\quad Re({\cal C}) \{
Re[\Delta(\alpha)\Delta(\beta)^{*}] Re({\cal S}) 
- Im[\Delta(\alpha)\Delta(\beta)^{*}] Im({\cal S}) \}
\nonumber\\
&& -Im({\cal C}) \{Im[\Delta(\alpha)\Delta(\beta)^{*}] Re({\cal S})+
Re[\Delta(\alpha)\Delta(\beta)^{*}] Im({\cal S}) 
\}.
\la{eq_cp1b}
\eeqn
For CP conserving interactions all couplings can be chosen real 
\ci{Bartl_nucl} and the last
two terms in (\re{eq_cp1b}) vanish. The second and third term is due to
interference between s--channel exchange of particles with finite width and
the crossed channels. Their contributions are noticeable  
only in the neighbourhood of the pole of the exchanged particles,
whereas far from the pole they are
proportional to the width of the exchanged particles and therefore 
negligible. In the second and fourth term $Im({\cal S})$ originates from
products of momenta and polarization
vectors with the Levy--Civita tensor $\epsilon_{\mu \nu \rho \sigma}$.
These two terms lead to triple product correlations of momenta
which are sensitive to CP--violation \ci{Christova}.
\subsection{CPT invariance}\la{sec:31}
Under CPT the helicity states of a Dirac fermion $f_D$ and a Majorana
fermion $f_M$, respectively, transform as
\beqn
&&|f_{D}(\vec{p},\lambda)> \stackrel{CPT}{\to}
|f_{D}(\vec{p},-\lambda)>, \\
&&|f_{M}(\vec{p},\lambda)> \stackrel{CPT}{\to}  
(-1)^{\lambda-\frac{1}{2}} \eta
|f_{M}(\vec{p},-\lambda)> \la{hans_1}.
\eeqn
Note that in addition to the CPT phase $\eta=\pm i$ for Majorana fermions
a helicity
dependent phase factor appears in (\re{hans_1}) which has observable 
consequences \ci{Kayser}.\\[.2em] 
{\it Majorana fermions:}\\
In the absence of absorptive effects, i.e. if the
finite width of the exchanged particles can be neglected, compare with 
(\ref{eq_cp1b}), 
the unitarity and CPT invariance of the S--matrix leads in 
lowest order  pertubation theory \cite{Petkov,Christova,Nagano}
to the following constraints for the production density matrix
of Majorana fermions:
\mathindent0cm
\bequ 
\rho_{P}^{\lambda_i \lambda_j, \lambda^{'}_i \lambda^{'}_j}
(P_{f}^{1,2,3}; P_{\bar{f}}^{1,2,3}; \Theta)=
\rho_{P}^{-\lambda_i -\lambda_j, -\lambda^{'}_i -\lambda^{'}_j}
(-P_{\bar{f}}^{1,2,3}; -P_{f}^{1,2,3}; \pi-\Theta)^{*}.
\la{hans_2}
\eequ
To derive (\ref{hans_2})
the CPT transformation has been supplemented
by a rotation ${\cal R}_2(\pi)$
around the normal to the production plane so that the beam direction
is unchanged, see Fig.~\ref{fig_1}a. 

Since we include beam polarization 
the spin density matrix of production (\re{eq4_4h}) also
depends on the polarization $P_{f}^{1,2,3}$, $P_{\bar{f}}^{1,2,3}$
of the initial beams $f$, $\bar{f}$ and is denoted by
$\rho^{\lambda_i \lambda_j, \lambda_i' \lambda_j'}_{P}(P_{f}^{1,2,3};
P_{\bar{f}}^{1,2,3};\Theta)$, where $\Theta$ denotes 
the angle between $f$ and $f_i$.
Note that in (\ref{hans_2}) 
the initial state polarizations of $f$ and $\bar{f}$ are
interchanged with a sign reversal.

Comparison with the expansion (\re{eq4_4h}) of $\rho_P$ 
 leads to the following classification of the different terms 
according to their CPT symmetry properties:
\begin{itemize}
\item Type S: $P$, $\Sigma_P^1$, $\Sigma_P^2$, $\Sigma_P^{11}$,
$\Sigma_P^{22}$, $\Sigma_P^{33}$, $\Sigma_P^{12}$, $\Sigma_P^{21}$ 
are symmetric,
\item Type A: $\Sigma_P^3$, $\Sigma_P^{13}$, $\Sigma_P^{31}$, 
$\Sigma_P^{23}$, $\Sigma_P^{32}$ are antisymmetric 
\end{itemize}
to the substitution 
$(P_{f}^{1,2,3}; P_{\bar{f}}^{1,2,3}; \Theta)\to 
(-P_{\bar{f}}^{1,2,3}; -P_{f}^{1,2,3}; \pi-\Theta)$. 

For unpolarized beams terms of Type S are FB--symmetric 
whereas the members of Type A are FB--antisymmetric. 
These symmetry properties hold also for production of Majorana fermions with
longitudinally polarized $e^+ e^-$ beams, where the dependence on the beam 
polarization is given by $(1-P_{e^-} P_{e^+})$ and
$(P_{e^+}-P_{e^-})$.

If the finite width is taken into account these symmetry
properties are violated and the terms obtain contributions
which are far from the resonances proportional to their widths and
negligible. Close to the resonances, however, these contributions may be 
noticeable.\\[.2em]
{\it Dirac fermions:}\\
For the case of Dirac fermions CPT followed by the rotation ${\cal R}_2(\pi)$
relates the spin density matrix 
$\rho_P^{\lambda_i\lambda_j,\lambda'_i\lambda'_j}$ of 
$f \bar{f}\to f_i(p_i,\lambda_i) \bar{f}_j(p_j,\lambda_j)$ to 
$\tilde{\rho}_P^{-\lambda_j-\lambda_i,-\lambda'_j-\lambda'_i}$
of $f \bar{f}\to f_j(p_i,-\lambda_j) \bar{f}_i(p_j,-\lambda_i)$.

We get in this case instead of (\ref{hans_2})
\mathindent0cm
\beqn 
&& \rho_{P}^{\lambda_i \lambda_j, \lambda^{'}_i \lambda^{'}_j}
(P_{f}^{1,2,3}; P_{\bar{f}}^{1,2,3}; \Theta)\nonumber\\
&&=
(-1)^{\lambda_i+\lambda'_i+\lambda_j+\lambda'_j}
\tilde{\rho}_{P}^{-\lambda_j -\lambda_i, -\lambda^{'}_j -\lambda^{'}_i}
(-P_{\bar{f}}^{1,2,3}; -P_{f}^{1,2,3}; \Theta)^{*},
\la{hans_7}
\eeqn
where $\Theta$ is the angle between the direction of beam particle
$f$ and the outgoing fermion $f_i$ and $f_j$, respectively.
For polarized beams the polarizations
of $f$ and $\bar{f}$ are interchanged with sign reversal.
However, in contrast to the Majorana case,
(\ref{hans_7}) does not result in constraints for the angular dependence.
\subsection{CP--invariance}\la{sec:32}
We study the consequences of CP invariance for  
the production of two Majorana fermions, as illustrated in Fig.~\ref{fig_1}b, 
or two Dirac fermions.
In this case all couplings can be chosen real \ci{Bartl_nucl}.

Under CP the helicity states of a Dirac and a Majorana fermion transform as
\beqn
&&|f_{D}(\vec{p},\lambda)> \stackrel{CP}{\to}
|\bar{f}_{D}(-\vec{p},-\lambda)>,\la{hans_6}\\
&&|f_{M}(\vec{p},\lambda)> \stackrel{CP}{\to}\eta^{CP} 
|f_{M}(-\vec{p},-\lambda)>,\quad\mbox{with }\eta^{CP}=\pm i.
\eeqn
{\it Majorana fermions:}\\
If CP invariance holds this leads for the joint production density matrix
to the constraints
\mathindent0cm
\bequ 
\rho_{P}^{\lambda_i \lambda_j, \lambda^{'}_i \lambda^{'}_j}
(P_{f}^{1,2,3}; P_{\bar{f}}^{1,2,3};\Theta)=
\rho_{P}^{-\lambda_i -\lambda_j, -\lambda^{'}_i -\lambda^{'}_j}
(P_{\bar{f}}^1,-P_{\bar{f}}^{2,3};P_{f}^1,-P_{f}^{2,3}; \pi-\Theta).
\la{hans_4}
\eequ
Then one obtains the following CP symmetry properties for the 
terms of $\rho_P$ (\ref{eq4_4h}):
\begin{itemize}
\item Type S': $P$, $\Sigma_P^1$, $\Sigma_P^{11}$,
$\Sigma_P^{22}$, $\Sigma_P^{33}$, $\Sigma_P^{23}$, $\Sigma_P^{32}$ 
are symmetric, 
\item Type A': $\Sigma_P^2$, $\Sigma_P^3$, $\Sigma_P^{12}$, $\Sigma_P^{21}$,
$\Sigma_P^{13}$, $\Sigma_P^{31}$
 are antisymmetric
\end{itemize}
to the substitution
$(P_{f}^{1,2,3}; P_{\bar{f}}^{1,2,3}; \Theta)\to 
(P_{\bar{f}}^1,-P_{\bar{f}}^{2,3}; P_{f}^1,-P_{f}^{2,3}; \pi-\Theta)$.

For unpolarized beams terms of Type S' are FB--symmetric 
whereas the members of Type A' are FB--antisymmetric. 
The FB--symmetry of the term $P$ reflects a well--known
property of the differential cross sections for unpolarized beams \ci{Petkov}.
These symmetry properties hold also for production of Majorana fermions with
longitudinally polarized $e^+ e^-$ beams, where the dependence on the beam 
polarization is given by $(1-P_{e^-} P_{e^+})$ and
$(P_{e^+}-P_{e^-})$.

If the widths of the particles exchanged in the s--channel are 
neglected so that the CPT symmetry properties (\re{hans_2}) hold, this 
leads to 
\bequ
\Sigma_P^2=0,\quad \Sigma_P^{12}=0=\Sigma_P^{21}, \quad 
\Sigma_P^{23}=0=\Sigma_P^{32}.
\la{hans_5}
\eequ
If the finite width is taken into account these terms obtain contributions
which are far from the resonances proportional to their widths and
negligible. Close to the resonances, however, these contributions may be 
noticeable.\\[.2em]
{\it Dirac fermions:}\\
For the case of Dirac fermions CP 
relates the spin density matrix 
$\rho_P^{\lambda_i\lambda_j,\lambda'_i\lambda'_j}$ of
$f \bar{f}\to f_i(p_i,\lambda_i) \bar{f}_j(p_j,\lambda_j)$ to 
$\tilde{\rho}_P^{-\lambda_j-\lambda_i,-\lambda'_j-\lambda'_i}$
of $f \bar{f}\to f_j(p_i,-\lambda_j) \bar{f}_i(p_j,-\lambda_i)$
and we get in this case instead of (\ref{hans_4})
\mathindent0cm
\bequ 
\rho_{P}^{\lambda_i \lambda_j, \lambda^{'}_i \lambda^{'}_j}
(P_{f}^{1,2,3}; P_{\bar{f}}^{1,2,3}; \Theta)=
\tilde{\rho}_{P}^{-\lambda_j -\lambda_i, -\lambda^{'}_j -\lambda^{'}_i}
(P_{\bar{f}}^1,-P_{\bar{f}}^{2,3};P_{f}^1,-P_{f}^{2,3}; \Theta),
\la{hans_8}
\eequ
where $\Theta$ is the angle between the direction of beam particle
$f$ and the outgoing fermion $f_i$ and $f_j$, respectively.
For polarized beams the polarizations $P^{2,3}$  
of $f$ and $\bar{f}$ are interchanged with sign reversal. 
In contrast to the Majorana case (\ref{hans_8}) 
does, however, not result in constraints for the angular dependence.
\subsubsection{CP--violation}\la{sec:321}
\bsl
If CP is violated, the couplings are complex  
and the terms of the second line in (\re{eq_cp1b}) are contributing. 
We distinguish the two types of CP--violating terms: 
\begin{itemize}
\item CP--violating terms of Type I with the structure\\ 
$Im({\cal C}) 
Re[\Delta(\alpha)\Delta(\beta)^{*}] Im({\cal S})$
\item CP--violating terms of Type II with the structure\\
$Im({\cal C}) Im[\Delta(\alpha)\Delta(\beta)^{*}] Re({\cal S})$.
\end{itemize}
{\it Majorana fermions:}\\
One can show that the CP--violating terms in (\re{eq_cp1b})
show opposite FB--dependence than the CP conserving ones:
terms of type $S'$ obtain now contributions which are FB--antisymmetric, 
whereas all termes of type $A'$ obtain contributions which are FB--symmetric.
This results in nonvanishing polarization $\Sigma_P^2$ perpendicular to the 
production plane and nonvanishing spin--spin correlations
$\Sigma_P^{12}$,$\Sigma_P^{21}\neq 0$ and 
$\Sigma_P^{23}$,$\Sigma_P^{32}\neq 0$.
\esl

\noindent Additional CP--violating terms:\\[-2em]
\begin{itemize}
\item Type I: 
 a) $\Sigma^2_P$, $\Sigma^{12}_P$, $\Sigma^{21}_P$ are
FB--symmetric\\
\phantom{Type I:} 
b) $\Sigma^{23}_P$, $\Sigma^{32}_P$ are FB--antisymmetric
\\[-1em]
\item Type II:
 a) $\Sigma^3_P$, $\Sigma^{13}_P$, $\Sigma^{31}_P$ are FB--symmetric\\
\phantom{Type II:} 
b) $P$, $\Sigma^1_P$, $\Sigma^{11}_P$, $\Sigma^{22}_P$,
                              $\Sigma^{33}_P$ are FB--antisymmetric.
\end{itemize} 
Note that in the case of CP--violation the
FB-symmetry of the $P$--terms
which is characteristic for production of Majorana fermions  \ci{Petkov}
is preserved if the width of the exchanged particles is neglected
since in this case 
the predictions of CPT--invariance holds.\\[.2em]
The CP and CPT symmetries for Majorana fermions are summarized in 
Table~\re{tab_-1}.
\section{Energy and angular distributions of the decay products}
\la{sec:40}
In this section we study for the three--particle decay into fermions
the influence of the polarization and of the 
spin--spin correlations of the decaying Majorana  or Dirac fermions
on 
the energy spectrum and angular distributions of the decay fermions in the
lab frame.
If the decay of only 
one of the produced fermions, e.g. $f_i$, is considered, one has to set 
in (\re{eq4_5}) $\Sigma^b_D(f_j)\equiv 0$ and $D(f_j)\equiv 1$.
 The total cross section for the complete process of
production and decay is given by 
$\sigma_P(f \bar{f}\to f_i f_j)\times BR(f_i\to f_{i1} f_{i2} f_{i3})$.
\subsection{Opening angle distribution and energy spectrum of the decay 
leptons}\la{sec:41}
For the distribution of the opening angle $\Theta_{i2,i3}$ between the 
fermions $f_{i2}$ and $f_{i3}$ from the decay 
$f_i\to f_{i1} f_{i2} f_{i3}$ it is favourable to parametrize the phase 
space of the decay products by the polar angle $\Theta_{i,i2}$ between the 
momenta of $f_{i}$ and $f_{i2}$, the azimuthal angle $\Phi_{i,i2}$, 
the opening angle $\Theta_{i2,i3}$ between the momenta of $f_{i2}$ and 
$f_{i3}$ and the corresponding 
azimuthal angle $\Phi_{i2,i3}$ (Fig.~\re{fig_2}):
\bequ
d\sigma= {\cal F}|T|^2 
\sin\Theta d\Theta d\Phi \sin\Theta_{i,i2} d\Theta_{i,i2}d\Phi_{i,i2}
 \sin\Theta_{i2,i3}d\Theta_{i2,i3}d\Phi_{i2,i3}dE_{i2}, \la{eq23_2a}
\eequ
where $\Theta$ is the production angle and 
\bequ
{\cal F}=\frac{q}{2^{11} (2\pi)^7  m_i \Gamma_{i}E_b^3}
\frac{E_{i2} [m_i^2-m_{i1}^2-2 E_{i2}(E_i-q\cos\Theta_{i,i2})]}
{[E_i-q\cos\Theta_{i,i3}-E_{i2}(1-\cos\Theta_{i2,i3})]^2}. \la{eq_F}
\eequ
$E_b$ denotes the beam energy,
$E_{\alpha}$ denotes the energy of the fermion $f_{\alpha}$ in the lab frame
and 
the momentum $q$ is given by 
$q=|\vec{p}_i|=|\vec{p}_j|=
\frac{\sqrt{\lambda(s,m_i^2, m_j^2)}}{2\sqrt{s}}$ with
$\lambda(x,y,z)=x^2+y^2+z^2-2xy-2xz-2yz$. 
All polar (azimuthal) angles are denoted by $\Theta_{\alpha \beta}$ 
($\Phi_{\alpha \beta}$), 
where the first index denotes the polar axis. 

For the energy distribution the phase space can be parametrized
by the polar angle $\Theta_{i,i3}$ between the momenta of 
$f_i$ and $f_{i3}$ and the azimuth $\Phi_{i,i3}$ of $f_{i3}$
instead of the opening angle $\Theta_{i2,i3}$
and the azimuth $\Phi_{i2,i3}$:
\bequ
d\sigma= {\cal F} |T|^2 
 \sin\Theta d\Theta d\Phi \sin\Theta_{i,i2} d\Theta_{i,i2}d\Phi_{i,i2}
 \sin\Theta_{i,i3}d\Theta_{i,i3}d\Phi_{i,i3}dE_{i2}.
\la{eq23_1}
\eequ
With both these parametrizations the phase space of production
and decay factorizes exactly in the phase space 
of production ($\sin\Theta d\Theta d\Phi$)
and that of the decay, which is independent of $\Theta$.

To study the influence of spin correlations we distinguish between
the contributions from transverse polarization and from longitudinal 
polarization of the decaying fermion.

\begin{itemize}
\item
Transverse Polarization \\
In the contributions of the transverse polarizations in (\re{eq4_5}),
the terms 
\bequ
\Sigma^{1,2}_D(f_i)=\tilde{\Sigma}^{1,2\mu}_D(f_i)
s^{1,2 \mu}(f_i) \la{eq_trans1}
\eequ
depend on the azimuth $\Phi_{i,i2}$ between the production plane and the 
plane defined by the decaying fermion $f_i$ and the decay product $f_{i2}$.
To separate the $\Phi_{i,i2}$ dependence we introduce a new system of 
polarization vectors $t^{a \nu}(f_i)$. Here
$t^{3 \nu}(f_i)=s^{3 \nu}(f_i)$, whereas in the lab system $\vec{t^2}(f_i)$ is 
perpendicular to the plane defined by the momenta of $f_{i}$ and 
$f_{i2}$ and $\vec{t^1}(f_1)$ is in the plane orthogonal to the 
momentum of the decaying fermion $f_i$. Thus in (\re{eq_trans1}) the 
transverse polarization vectors $s^{1,2 \nu}(f_i)$ are
\beqn
s^{1\nu}(f_i)&=&
\cos\Phi_{i,i2} t^{1\nu}(f_i)-\sin\Phi_{i,i2} t^{2\nu}(f_i),
\la{eq3_1c}\\
s^{2\nu}(f_i)&=&
\sin\Phi_{i,i2} t^{1\nu}(f_i)+\cos\Phi_{i,i2} t^{2\nu}(f_i).
\la{eq3_1d}
\eeqn
{\it Majorana and Dirac fermions:}\\
{\it Since the phase space parametrizations
(\ref{eq23_2a})--(\ref{eq23_1}) are independent of $\Phi_{i,i2}$ 
the contributions of 
$\Sigma^{1,2}_D(f_i)$ to the opening angle distribution and to the lepton 
energy spectrum vanish for both Majorana and Dirac fermions 
due to the integration over $\Phi_{i,i2}$ \ci{Gudi_diss}.}

\item
Longitudinal Polarization\\
{\it Majorana fermions:}\\
For Majorana fermions the longitudinal polarization $\Sigma^3_P(f_i)$ is 
forward--backward antisymmetric if CP is conserved, see section \ref{sec:32}. 
Due to the 
factorization of the phase space in production and decay
also the contribution of the 
longitudinal polarization vanishes after integration over 
the production angle $\Theta$.

{\it Consequently both the
energy and the opening angle distribution of the decay products
of Majorana fermions in the laboratory system 
are independent of spin correlations and factorizes exactly
in production and decay if CP is conserved.} 

If, however, CP is violated and if the width
of the exchanged particle has to be taken into account, see (\ref{eq_cp1b}),
this factorization of the energy distribution and of the  
opening angle distribution of the decay products 
of Majorana fermions is violated,
since the longitudinal
polarization $\Sigma^3_P$ gets FB--symmetric
contributions from CP--violating terms.
Since these additional terms are of Type II,
they are for energies far from the resonance
proportional to the width of the exchanged particle.\\[.2em]
{\it Dirac fermions:}\\
{\it The longitudinal polarization  
of produced Dirac fermions
is not forward--backward asymmetric.
It influences the 
energy spectra and opening angle distributions of the decay products so that
they do not factorize in production and decay.}
\end{itemize}
\subsection{Decay lepton angular distribution}
\la{sec:4}
The decay lepton angle $\Theta_{1,i2}$, see Fig.~\re{fig_2},
denotes the angle between the incoming fermion $f$ and
the outgoing fermion $f_{i2}$ from the decay
$f_i \to f_{i1} f_{i2} f_{i3}$.
Since the decay angular distribution of $f_i$ 
depends on its polarization $\Sigma^a_P$,
 the angular distribution $d\sigma/d\cos\Theta_{1,i2}$ in the lab frame is
sensitive to the spin correlations between production and decay. 
In this case we parametrize the 
phase space of production and decay by
\bequ
d\sigma = {\cal F} |T|^2 
\sin\Theta d\Theta d\Phi \sin\Theta_{1,i2} d\Theta_{1,i2}d\Phi_{1,i2}
 \sin\Theta_{1,i3}d\Theta_{1,i3}d\Phi_{1,i3}dE_{i2}.\la{eq23_3a}
\eequ

With this parametrization one has to express in ${\cal F}$ (\re{eq_F}) the
angles $\Theta_{i,i2}$, $\Theta_{i,i3}$, $\Theta_{i2,i3}$ 
and in (\re{eq3_1c}), (\re{eq3_1d}) the azimuth $\Phi_{i,i2}$ by
the decay lepton angle $\Theta_{1,i2}$, the production angle $\Theta$ and the
azimuth angles $\Phi_{1,i2}$ and $\Phi$ \ci{Gudi_diss}.
Then neither the contributions of
the transverse polarizations, $\Sigma_D^{1,2}(f_i)$,
nor that of the longitudinal polarization, $\Sigma_P^3(f_i)$, vanish 
due to phase space integration.\\[.2em]
{\it Majorana and Dirac fermions:}\\
{\it Consequently neither for Dirac fermions  
nor for Majorana fermions the decay
lepton angular distribution in the lab frame factorizes in production
and decay but depend sensitively on spin correlations.}
\subsection{Siamese opening angle distribution}
\la{sec:43}
The siamese opening angle  $\Theta_{j2,i2}$
denotes the angle between decay products $f_{i2}$ and
$f_{j2}$ from the decay of different particles $f_i$ and $f_j$.
Since the angular distributions of the decay products
depend on the polarizations of the
mother particles $f_i$, $f_j$, the distribution $d\sigma/d\cos\Theta_{j2,i2}$
of the opening angle between them is determined by the spin--spin
correlations $\Sigma_P^{ab}$, see (\ref{eq4_5}),
between the decaying fermions. For the siamese
opening angle distribution it is favourable to parametrize the phase space for 
production and decay by
\beqn
d\sigma&=& {\cal F}{\cal G}|T|^2 
\sin\Theta d\Theta d\Phi\sin\Theta_{i, j2} d\Theta_{i,j2}d\Phi_{i,j2}
 \sin\Theta_{i,j3}d\Theta_{i,j3}d\Phi_{i,j3}dE_{j2}\nonumber\\
&&\times \sin\Theta_{i,i3} d\Theta_{i,i3}d\Phi_{i,i3}
\sin\Theta_{j2,i2} d\Theta_{j2,i2}d\Phi_{j2,i2} dE_{i2},
\la{eq34_5}
\eeqn
where ${\cal F}$ is given by (\re{eq_F}) and ${\cal G}$ is 
given by 
\beqn
{\cal G}&=&\frac{1}{2^5 (2 \pi)^5 m_j \Gamma_j}
\frac{E_{j2} [m_j^2-m_{j1}^2-2 E_{j2}(E_j-q\cos\Theta_{i,j2})]}
{[E_j-|\vec{p}_j|\cos\Theta_{i,j3}-E_{j2}(1-\cos\Theta_{j2,j3})]^2}. \la{eq_G}
\eeqn
Note that in (\ref{eq34_5}), (\ref{eq_G}) also the angles 
$\Theta_{i,j2}$, $\Phi_{i,j2}$, $\Theta_{i,j3}$, $\Phi_{i,j3}$
are given with respect to the direction of $f_i$.
 
With the same arguments as for the contributions from transverse 
polarization to the opening angle distribution one infers that
in the case of CP--conservation
only the diagonal terms $\Sigma_P^{11}$, $\Sigma_P^{22}$ and
$\Sigma_P^{33}$ contribute for both Majorana and Dirac fermions.
The analytical relations between all angles 
used for the different parametrizations
 are explicitly given in \ci{Gudi_diss}.
The total cross section is given
by $\sigma_P(f \bar{f}\to f_i f_j)\times BR(f_i)\times BR(f_j)$.\\[.2em]
{\it Majorana and Dirac fermions:}\\
{\it The siamese opening angle distribution 
factorizes neither for Majorana nor for Dirac 
fermions but depends on spin--spin correlations.} 
\section{Numerical results}
\la{sec:5}
In the following we give numerical examples for the influence of
spin correlations on the energy and angular distributions of leptons from
production and leptonic decay of
charginos as an example for Dirac fermions
\bequ
e^+e^-\to \tilde{\chi}^+_1 \tilde{\chi}^-_1,\quad
\tilde{\chi}^-_1\to \tilde{\chi}^0_1 e^- \bar{\nu}_e, \la{eq5_2}
\eequ
and of neutralinos as an example for Majorana fermions,
\bequ
e^+ e^-\to \tilde{\chi}^0_1 \tilde{\chi}^0_2,\quad 
\tilde{\chi}^0_2\to \tilde{\chi}^0_1 e^+ e^-. \la{eq5_1}
\eequ
We choose a CMSSM scenario \ci{Blair} 
with the gaugino mass parameters $M_1=78.7$~GeV and $M_2=152$~GeV,
the higgsino mass parameter $\mu=316$~GeV and the ratio of the higgs 
expectation values $v_2/v_1=\tan\beta=3$.
Then $\tilde{\chi}^0_{1,2}$ and $\tilde{\chi}^{\pm}_1$ are gaugino--like
with masses $m_{\tilde{\chi}^0_1}=71$~GeV,
$m_{\tilde{\chi}^0_2}=130$~GeV and $m_{\tilde{\chi}^\pm_1}=128$~GeV.
The slepton masses are $m_{\tilde{e}_L}=176$~GeV, $m_{\tilde{\nu}}=161$~GeV, 
$m_{\tilde{e}_R}=132$~GeV.
The total widths $\Gamma_{\tilde{\chi}^{\pm}_1}$,
$\Gamma_{\tilde{\chi}^0_2}$ are
of O(keV), so that the narrow width
approximation is well justified.

Here we study the CP--conserving case and
give numerical results for $e^+e^-$--cms
energies close to threshold and at $\sqrt{s}=500$~GeV for
unpolarized and for longitudinally polarized beams with
$P_{e^-}=\pm 85\%$ and $P_{e^+}=\mp 60\%$.
The  total cross sections are listed in Table~\re{tab_0}.
\subsection{Distributions in chargino production and decay }
\la{sec:51}
{\it a) Lepton energy und opening angle distribution}\\
In both distributions only 
the longitudinal polarization $\Sigma^3_P$ of the chargino
contributes. We show the energy spectrum of the decay electron
for $\sqrt{s}=270$~GeV in 
Fig.~\re{fig_3}a, and for $\sqrt{s}=500$~GeV in Fig.~\re{fig_3}b.
For both energies and for all polarizations the spin correlations 
flatten the energy spectrum. Their influence is strongest near
the maximum of the spectrum where 
neglecting $\Sigma^3_P(\tilde{\chi}^{\pm}_1)$ could result in
higher cross sections of about 5\% for $\sqrt{s}=270$~GeV and even 20\%
for $\sqrt{s}=500$~GeV.

Since the opening angle distribution is unobservable in the leptonic decay we
show no figures. It can, however, be observed as the opening
angle between the jet axes in the hadronic decay
$\tilde{\chi}^-_1\to \tilde{\chi}^0_1 \bar{u} d$.
Depending on the masses of the exchanged
squarks the influence of spin correlations amounts up to 20\% 
\ci{Gudi_talk}.\\[.5em]
{\it b) Decay angular distribution}\\
In the angular distribution of the decay $e^-$ 
the influence of spin correlations is highest close to threshold. 
All polarizations $\Sigma_P^1$, $\Sigma_P^2$ and $\Sigma_P^3$ are 
contributing.

In the case of 
$e^+ e^-\to \tilde{\chi}^+_1 \tilde{\chi}^-_1$, 
$\tilde{\chi}^-_1\to \tilde{\chi}^0_1 e^- \bar{\nu}$ and
for $\sqrt{s}=270$~GeV
the forward--backward asymmetry is $A_{FB}=33\%$
in our scenario, Fig.~\ref{fig_4}a. Neglecting the effects of 
chargino polarizations would lead to $A_{FB}$ of only $3\%$! This 
demonstrates impressively the importance of spin correlations.\\[.5em]
{\it c) Siamese opening angle distribution}\\
Fig.~\ref{fig_4}b shows
for $\sqrt{s}=270$~GeV the distribution of the opening angle between
the $e^+$ and $e^-$ from the
leptonic decay of both charginos $\tilde{\chi}_1^{\pm}$.

One can see that neglecting the spin correlations would result
in a more isotropic distribution. This can be illustrated
by the `parallel--antiparallel' asymmetry 
\bequ
A_{PA}=\frac{\sigma(\cos\Theta_{e^+e^-}>0)-\sigma(\cos\Theta_{e^+e^-}<0)}
{\sigma(\cos\Theta_{e^+e^-}>0)+\sigma(\cos\Theta_{e^+e^-}<0)}.
\la{eq_PA}
\eequ
For $\sqrt{s}=270$~GeV  one obtains $A_{PA}=-10\%$ for all beam polarizations
whereas neglecting spin--spin correlations 
would result in $A_{PA}\sim -5\%$.
\subsection{Distributions in neutralino production and decay}\la{sec:52}
{\it a) Lepton energy and opening angle distribution}\\
Due to the Majorana character of the neutralinos 
both distributions are independent of spin correlations between production 
and decay and will not further be discussed here \ci{Gudi_neut}.
As an example we show the opening angle distribution in our scenario for  
$\sqrt{s}=230$~GeV, see Fig.~\re{fig_5}a. Beam polarization
changes only the size of the distribution and has no
influence on the shape. \\[.5em]
{\it b) Decay angular distribution}\\
The influence of spin correlations is for neutralino production and
decay much more significant than for chargino production and decay. If one 
neglects spin correlations the lepton angular distribution would be
forward--backward symmetric as the production cross section due
to the Majorana character of the neutralinos. The influence of spin 
correlations, however, results in our scenario for
$\sqrt{s}=230$~GeV and $P_{e^-}=-85\%$, $P_{e^+}=+60\%$ in a significant
forward--backward asymmetry of $A_{FB}=-12\%$,
Fig.~\ref{fig_5}b.

The polarizations contributing to the energy and the different angular
distributions as well as the effect of spin correlations
for our examples are listed in Table~\re{tab_1}.
\section{Conclusion}
\la{sec:7}
We have studied the significance of spin correlations for
production  and subsequent decay of Majorana and Dirac
fermions in fermion--antifermion annihilation with polarized beams.
Crucial for the dependence on spin correlations
of the energy and angular distribution of the
decay particles is the forward--backward symmetry
or asymmetry of the various terms of the production spin--density matrix
derived from CPT-- and CP--invariance.
Decay angular and siamese opening angle distribution depend sensitively on
the polarization of the decaying particle for both Majorana and Dirac fermions.
In our numerical
examples for production and leptonic decay of neutralinos and charginos
neglecting spin correlations would result in 
a forward--backward asymmetry of the decay leptons which is 
one order of magnitude too small.
This demonstrates that the consideration of spin correlations is in general
indispensable for production and decay processes of spinning particles.

Decay energy and opening angle distributions 
of Majorana fermions are
exactly inpendent of the spin correlations between production and decay
if CP is conserved. This shows the possibility to
establish the Majorana character by comparing the 
measured decay energy and opening angle distributions with MC--studies
including the complete spin correlations.
\section{Acknowledgement}
\label{ack}
We thank A.~Bartl, E.~Christova and W.~Majerotto for valuable discussions.
This work was supported by the Bundesministerium f\"ur Bildung
und Forschung, contract No.\ 05 7WZ91P (0), by DFG FR 1064/4--1 and by
the Fonds zur F\"orderung der wissenschaftlichen Forschung of Austria,
project No.\ P13139-PHY.

\begin{table}
\hspace{-1cm}
\begin{tabular}{|l|lc|lc|}
\hline
 && $Im[\Delta(\alpha)\Delta(\beta)^{*}]=0$& 
&$Im[\Delta(\alpha)\Delta(\beta)^{*}]\neq 0$ \\ \hline
 CP & s: & $P$, $\Sigma^1_P$, $\Sigma^{11}_P$, $\Sigma^{22}_P$,
$\Sigma^{33}_P$ &s: &
$P$, $\Sigma^1_P$, $\Sigma^{11}_P$, $\Sigma^{22}_P$,
$\Sigma^{33}_P$, $\Sigma^{23}_P$, $\Sigma^{32}_P$\\
 & a: & $\Sigma^3_P$, $\Sigma^{13}_P$, $\Sigma^{31}_P$ & 
a: & $\Sigma^2_P$, $\Sigma^3_P$, 
$\Sigma^{12}_P$, $\Sigma^{21}_P$, $\Sigma^{13}_P$, $\Sigma^{31}_P$\\
 & \multicolumn{2}{c|}{$\Sigma^2_P=0$ , $\Sigma^{12}_P=0=\Sigma^{21}_P$, 
$\Sigma^{23}_P=0=\Sigma^{32}_P$}
&&\\ \hline
&&&\multicolumn{2}{c|}{Additional Contributions} \\
\mbox{$\not\!\!\!\!{\mbox{C}\mbox{P}}$} 
& s: & $P$, $\Sigma^1_P$, $\Sigma^2_P$, $\Sigma^{11}_P$, 
$\Sigma^{22}_P$, $\Sigma^{33}_P$, $\Sigma^{12}_P$, $\Sigma^{21}_P$ & 
s:& $\Sigma^3_P$, $\Sigma^{13}_P$, 
$\Sigma^{31}_P$\\
 & a: & $\Sigma^3_P$, $\Sigma^{13}_P$, $\Sigma^{31}_P$, 
$\Sigma^{23}_P$, $\Sigma^{32}_P$ & a: & $P$, $\Sigma^1_P$,   
$\Sigma^{11}_P$, $\Sigma^{22}_P$, $\Sigma^{33}_P$ \\
\hline
\end{tabular}
\caption{ 
The forward--backward symmetry (s) and --antisymmetry (a)
of all terms of the production spin--density 
matrix (\re{eq4_4h}) for Majorana fermions 
with and without consideration of the total width of the 
exchanged particles for CP--conservation (CP) or CP--violation 
( \mbox{$\not\!\!\!\!{\mbox{C}\mbox{P}}$}).  
\label{tab_-1}}
\vspace{.9cm}
\begin{tabular}{|l|c||c|c|c|}
\hline
& $\sqrt{s}$/GeV &
$(00)$ & $(-+)$ & $(+-)$  \\ \hline
$\sigma(e^+ e^-\to \tilde{\chi}^+_1 \tilde{\chi}^-_1)\times$
$BR(\tilde{\chi}^+_1\to \tilde{\chi}^0_1 e^+ \bar{\nu}_e)$/fb & 270 & 20 
& 60 & 1\\ 
 & 500 & 46 & 137 & 3 \\ \hline
$\sigma(e^+ e^-\to \tilde{\chi}^+_1 \tilde{\chi}^-_1)\times$
$BR(\tilde{\chi}^+_1\to \tilde{\chi}^0_1 e^+ \bar{\nu}_e)$& & & &\\
$\times BR(\tilde{\chi}^-_1\to \tilde{\chi}^0_1 e^- \nu_e)$/fb 
& 270 & 3 & 8 & 0.2 \\ \hline
$\sigma(e^+ e^-\to \tilde{\chi}^0_1 \tilde{\chi}^0_2)\times$
$BR(\tilde{\chi}^0_2\to \tilde{\chi}^0_1 e^+ e^-)$/fb & 230 & 6 & 10 & 7\\ 
\hline
\end{tabular}
\caption{Cross sections for production and decay 
with different beam polarizations in our reference scenario. 
The different configurations: 
unpolarized beams and $P_{e^-}=\mp 85\%$, $P_{e^+}=\pm 60\%$ are denoted
by $(00)$ and $(-+)$, $(+-)$.
\label{tab_0}}
\vspace{.6cm}
\hspace*{-.8cm}
\begin{tabular}{|l|cc|c||cc|c|}\hline
 & \multicolumn{3}{c||}{{Polarization of Dirac fermions}} & 
\multicolumn{3}{c|}{{Polarization of Majorana fermions}}\\
& \multicolumn{2}{|c|}{CP}  & \mbox{$\not\!\!\!\!{\mbox{C}\mbox{P}}$} 
& \multicolumn{2}{|c|}{CP} & \mbox{$\not\!\!\!\!{\mbox{C}\mbox{P}}$} \\ 
 Decay distrib.& Terms & Size & Terms &Terms & Size & Terms\\ \hline
{energy } & $\Sigma^{3}_{P}$ &  20\% &
$\Sigma^{3}_{P}$& none &  &
$\Sigma^{3}_{P}$\\
{opening angle} & $\Sigma^{3}_{P}$ 
& 20\% &
$\Sigma^{3}_{P}$& none & &
$\Sigma^{3}_{P}$\\
{angular} & $\Sigma^{1,2,3}_{P}$
 & {\scriptsize factor 10}& 
$\Sigma^{1,2,3}_{P}$ & 
$\Sigma^{1,2,3}_{P}$ & {\scriptsize $A_{FB}$ up to 12\%}
& $\Sigma^{1,2,3}_{P}$ \\
{siamese angle} & $\Sigma^{11,22,33}_{P}$ 
&{\scriptsize  factor 2} & $\Sigma^{ab}_{P}$ & 
$\Sigma^{11,22,33}_{P}$ 
& {\scriptsize factor 2}& 
$\Sigma^{ab}_{P}$  \\ \hline
\end{tabular}
\caption{For the energy und different angular distributions 
the contributing polarizations are specified. 
The polarization dependence is different 
for Majorana and Dirac fermions. Also denoted is the effect of 
the spin correlations in our reference scenario. 
If CP is violated all distributions
depend on spin correlations.   }
\label{tab_1}       
\end{table}

\begin{figure}
\begin{center}
{\setlength{\unitlength}{1cm}
\begin{picture}(12,5)
\put(-.3,-1.2){\includegraphics{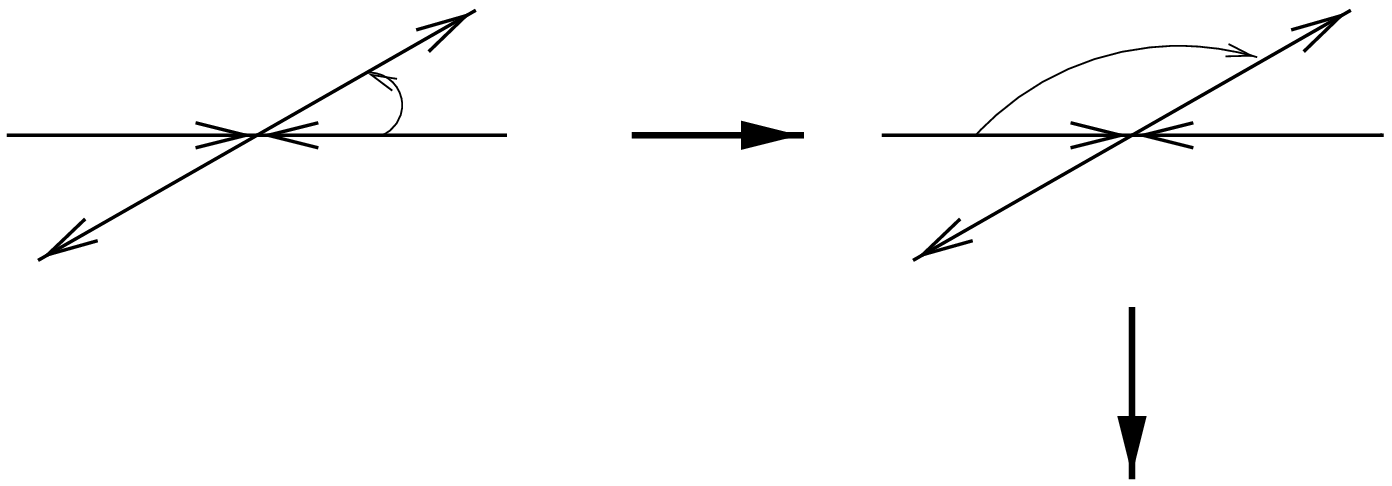}}
\put(-.5,2.5){\small$f(\vec{p}_1,\lambda_1)$}
\put(5.6,2.7){CPT}
\put(3.4,1.9){\small$\bar{f}(\vec{p}_2,\lambda_2)$}
\put(2.8,2.5){\small$\Theta$}
\put(1.8,3.7){\small $f_{M,i}(\vec{p}_i,\lambda_i)$}
\put(.9,1.1){\small $f_{M,j}(\vec{p}_j,\lambda_j)$}
\put(11,2.5){\small$f(-\vec{p}_1,-\lambda_2)$}
\put(7.1,1.9){\small$\bar{f}(-\vec{p}_2,-\lambda_1)$}
\put(9.4,2.7){\small$\pi-\Theta$}
\put(9.8,3.7){\small $f_{M,i}(\vec{p}_i,-\lambda_i)$}
\put(8.9,1.1){\small $f_{M,j}(\vec{p}_j,-\lambda_j)$}
\put(8.9,-.2){\small ${\cal R}_2(\pi)$}
\put(-.5,4){\small a)}
\end{picture}}\par\vspace{-1cm}
{\setlength{\unitlength}{1cm}
\begin{picture}(12,5)
\put(-.3,1){\includegraphics{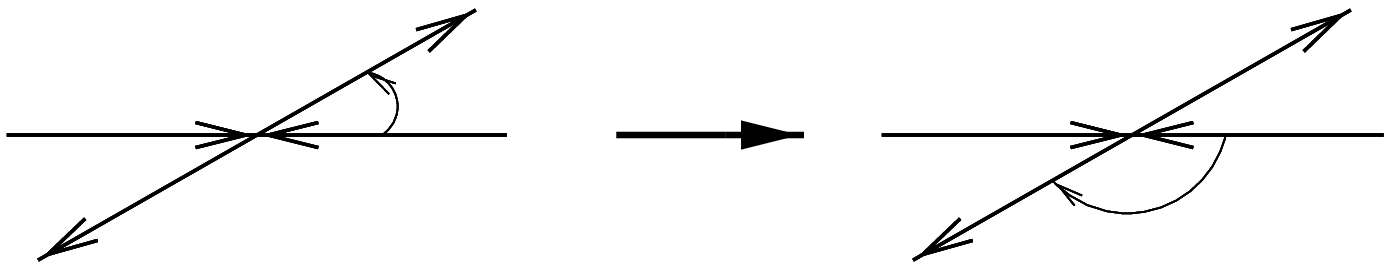}}
\put(-.5,2.5){\small$f(\vec{p}_1,\lambda_1)$}
\put(5.8,2.7){CP}
\put(3.4,1.9){\small$\bar{f}(\vec{p}_2,\lambda_2)$}
\put(2.8,2.5){\small$\Theta$}
\put(1.9,3.6){\small $f_{M,i}(\vec{p}_i,\lambda_i)$}
\put(.6,.9){\small $f_{M,j}(\vec{p}_j,\lambda_j)$}
\put(11.2,2.5){\small$\bar{f}(\vec{p}_2,-\lambda_1)$}
\put(7.2,1.9){\small$f(\vec{p}_1,-\lambda_2)$}
\put(9.6,1.8){\small$\pi-\Theta$}
\put(10.6,3.8){\small $f_{M,j}(-\vec{p}_j,-\lambda_j)$}
\put(8.6,.9){\small $f_{M,i}(-\vec{p}_i,-\lambda_i)$}
\put(-.5,4){\small b)}
\end{picture}}\par\vspace{-1.2cm}
\end{center}
\caption{Production of Majorana fermions under a) CPT
followed by a rotation ${\cal R}_2(\pi)$
and b) CP transformation. 
} \la{fig_1}
\vspace{1.5cm}
\begin{picture}(10,5)
\put(-.9,-.5){\includegraphics{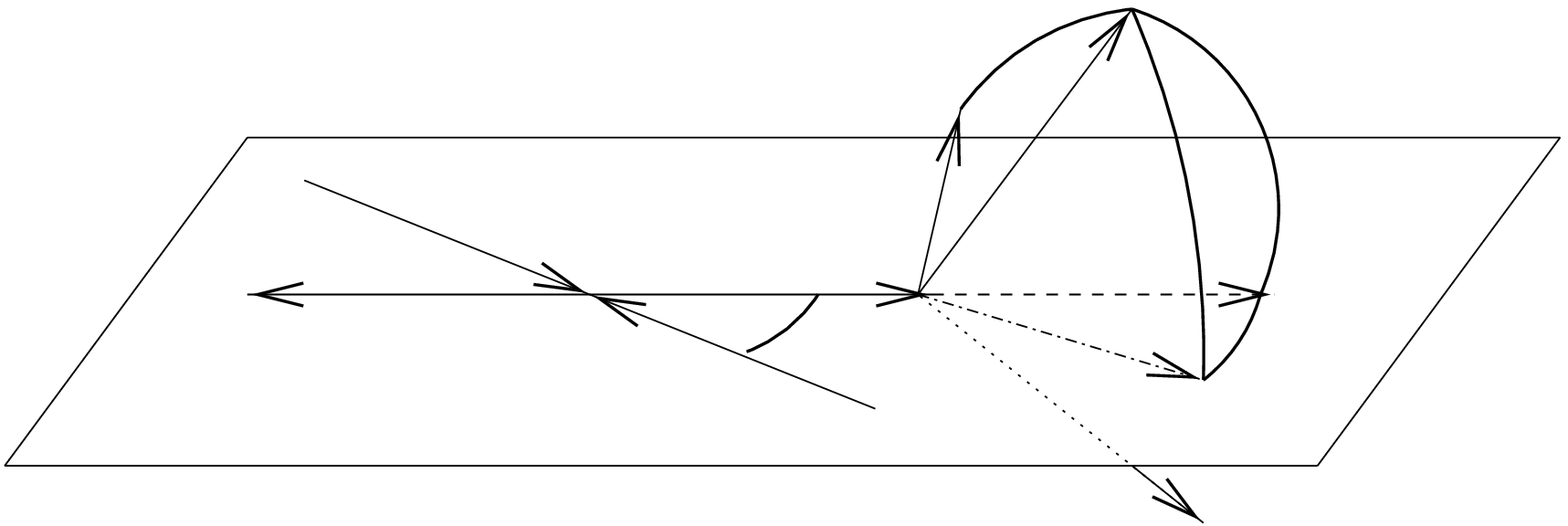}}
\put(2.6,3.4){ $\vec{p}_1$}
\put(8,.7){\small $\vec{p}_2$}
\put(7.7,2.4){\small $\vec{p}_i$}
\put(1.8,2.4){\small $\vec{p}_j$}
\put(7.1,1.5){\small $\Theta$}
\put(11.7,1.3){\small $\Theta$}
\put(11.3,.8){\small $\vec{p}_1$}
\put(7.9,3.5){\small $\vec{p}_{i3}$}
\put(10.4,5.5){\small $\vec{p}_{i2}$}
\put(10.1,-.3){\small $\vec{p}_{i1}$}
\put(8.5,5){\small $\Theta_{i2,i3}$}
\put(12,3.4){\small $\Theta_{i,i2}$}
\put(10.1,3){\small $\Theta_{1,i2}$}
\end{picture}\par\vspace{.2cm}
\caption{Definition of momenta and polar angles in the lab system.
The indices of the angles denote the plane covered by the corresponding 
momenta, the first index denotes the corresponding polar axis. 
The momenta and angles in the decay of $f_j$ are chosen
analogouesly.}
\label{fig_2}       
\end{figure}

\begin{figure}[t]
\hspace{-.9cm}
\begin{minipage}{7cm}
\begin{picture}(7,5)
\put(-.2,0){\includegraphics{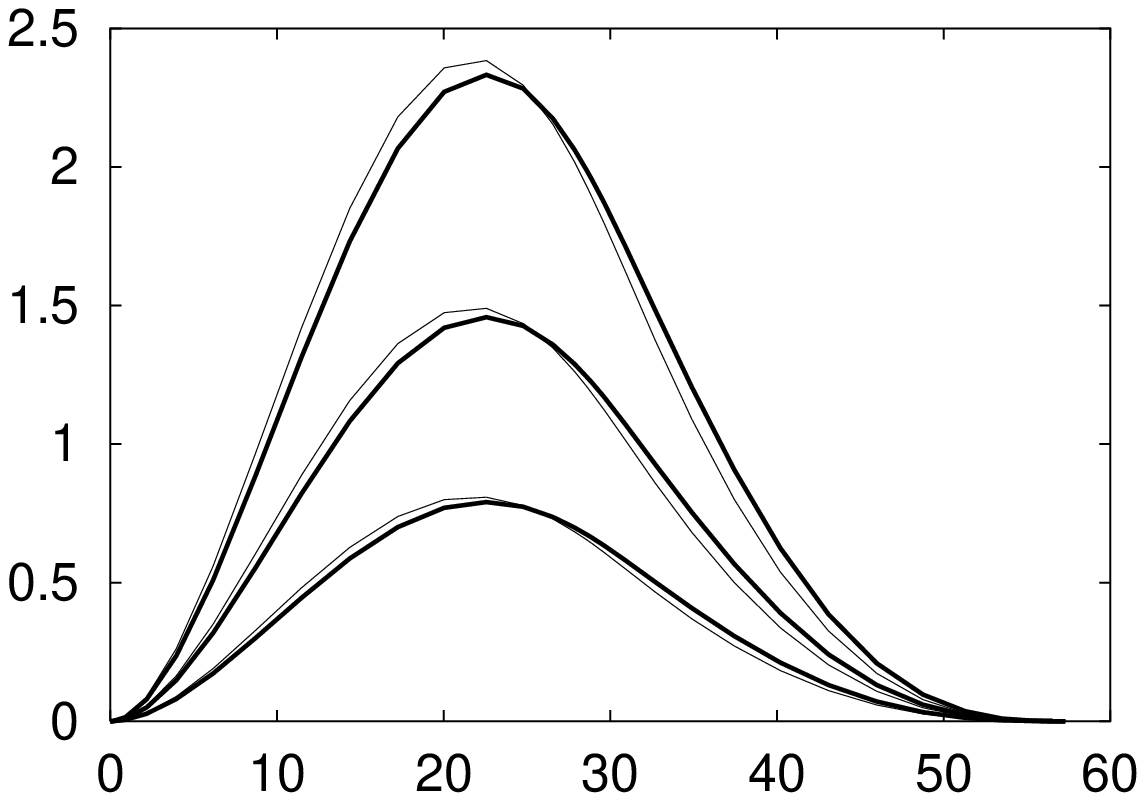}}
\put(4,4.2){\small $\sqrt{s}=270$~GeV}
\put(5.5,-.3){\small $ E_{e^-}/${\small ~GeV}}
\put(0,5.2){\small $d\sigma/dE_{e^-}$[fb/GeV]}
\put(2.5,1.5){\small $(+-)$}
\put(2.5,2.6){\small $(00)$}
\put(2.5,3.8){\small $(-+)$}
\put(3.5,5.2){\small a)}
\end{picture}\par\vspace{1cm}
\end{minipage}\hfill\hspace{.2cm}
\begin{minipage}{7cm}
\begin{picture}(7,5)
\put(-.2,0){\includegraphics{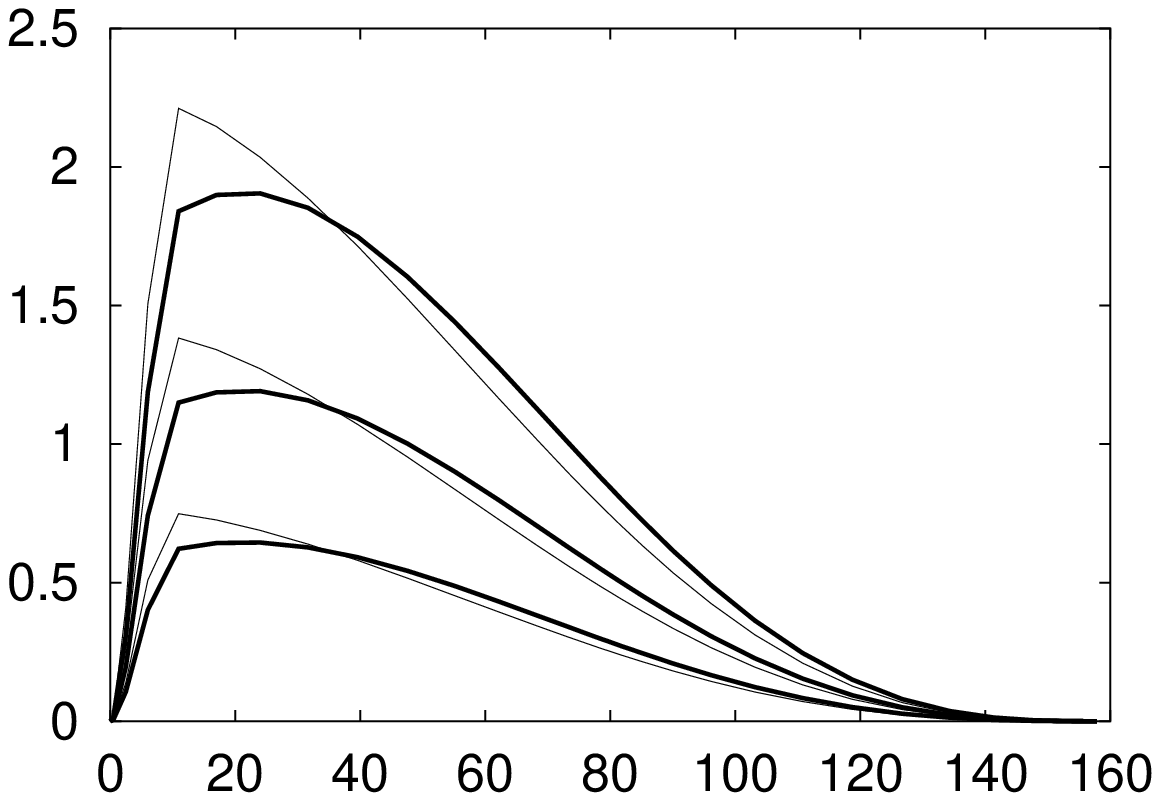}}
\put(4,4.2){\small $\sqrt{s}=500$~GeV}
\put(5.5,-.3){\small $ E_{e^-}/${\small ~GeV}}
\put(0,5.2){\small $d\sigma/dE_{e^-}$[fb/GeV]}
\put(2,1.7){\small $(+-)$}
\put(2,2.7){\small $(00)$}
\put(2.,3.8){\small $(-+)$}
\put(3.5,5.2){\small b)}
\end{picture}\par\hspace{-.5cm}\vspace{+.6cm}
\end{minipage}
\vspace{-1cm}
\caption{$e^+ e^-\to \tilde{\chi}^+_1 \tilde{\chi}^-_1$, 
$\tilde{\chi}^-_1\to \tilde{\chi}^0_1 e^- \bar{\nu}_e$: Energy distribution 
of decay $e^-$ a) at $\sqrt{s}=270$~GeV and b) at $\sqrt{s}=500$~GeV
for unpolarized beams (00) and both 
beams polarized $P_{e^-}=\mp 85\%$, $P_{e^+}=\pm 60\%$, $(-+)$ and $(+-)$,
with (thick lined) and without (thin lined) spin correlations
between production and decay. Close to the maximum 
the effect of chargino polarization is about in a) 5\% and in b) 20\%.
\la{fig_3}}
\vspace{.9cm}
\hspace{-.9cm}
\begin{minipage}{7cm}
\begin{picture}(7,5)
\put(-.2,0){\includegraphics{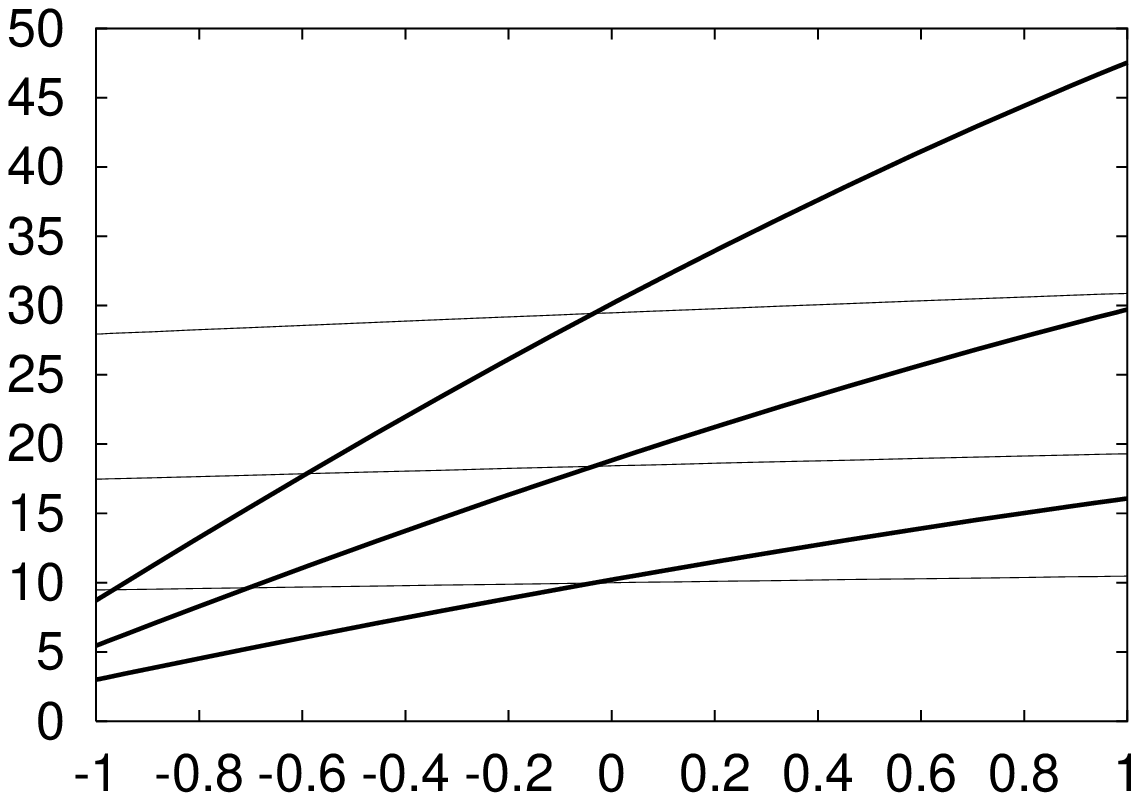}}
\put(1,4.2){\small $\sqrt{s}=270$~GeV}
\put(5.5,-.2){\small $\cos\Theta_{e^-}$}
\put(0,5.2){\small $d\sigma/d\cos\Theta_{e^-}$[fb]}
\put(3.5,5.2){\small a)}
\put(3,1.7){\small $(+-)$}
\put(3.7,2.7){\small $(00)$}
\put(3.9,4){\small $(-+)$}
\end{picture}\par\vspace{.3cm}
\end{minipage}\hfill\hspace{.2cm}
\begin{minipage}{7cm}
\begin{picture}(7,5)
\put(-.2,0){\includegraphics{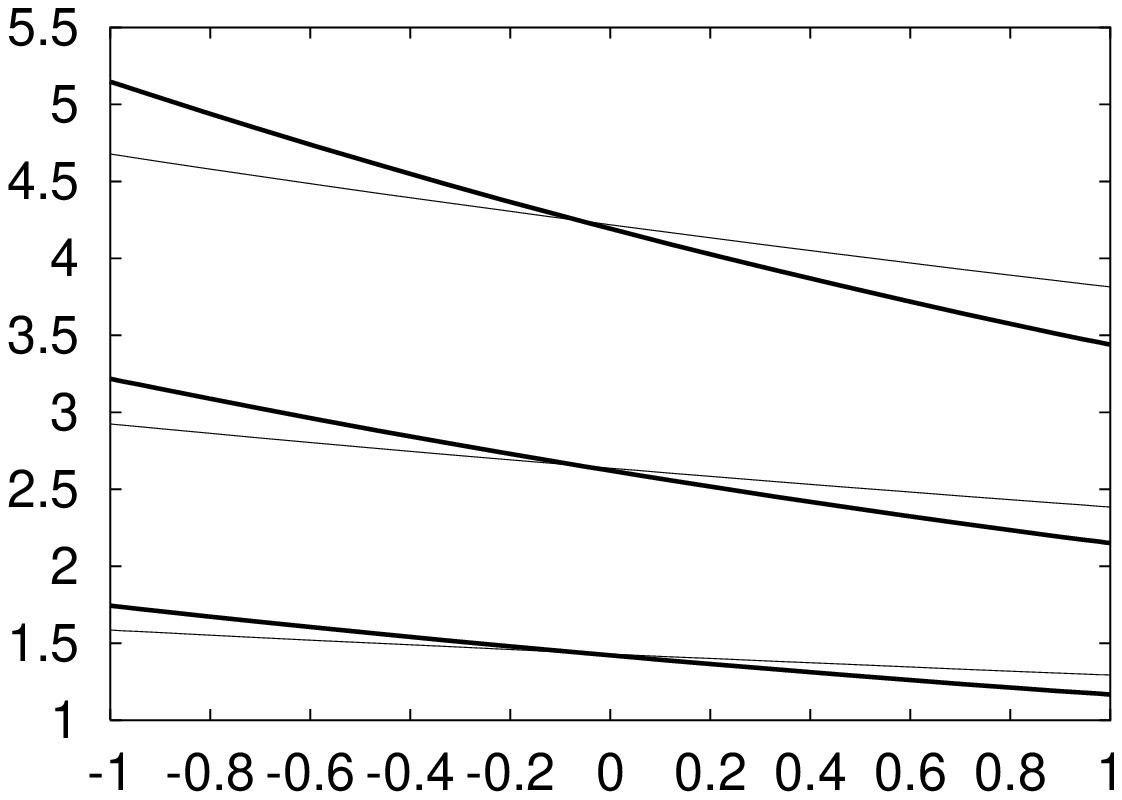}}
\put(4,4.2){\small $\sqrt{s}=270$~GeV}
\put(5.5,-.2){\small $\cos\Theta_{e^+e^-}$}
\put(0,5.2){\small $d\sigma/d\cos\Theta_{e^+e^-}$[fb]}
\put(3.5,5.2){\small b)}
\put(2.7,1.3){\small $(+-)$}
\put(2.4,2.6){\small $(00)$}
\put(2,4.3){\small $(-+)$}
\end{picture}\par\vspace{.3cm}
\end{minipage}
\caption{Chargino production and decay:
a) angular distribution 
of the decay $e^-$ in
$e^+ e^-\to \tilde{\chi}^+_1 \tilde{\chi}^-_1$, 
$\tilde{\chi}^-_1\to \tilde{\chi}^0_1 e^- \bar{\nu}_e$
and b) siamese  opening angle distribution of the decay $e^+$ and $e^-$ in
$e^+ e^-\to \tilde{\chi}^+_1 \tilde{\chi}^-_1$, 
$\tilde{\chi}^+_1\to \tilde{\chi}^0_1 e^+ \nu_e$,
$\tilde{\chi}^-_1\to \tilde{\chi}^0_1 e^- \bar{\nu}_e$ 
at $\sqrt{s}=270$~GeV 
for unpolarized beams (00) and both 
beams polarized $P_{e^-}=\mp 85\%$, $P_{e^+}=\pm 60\%$, $(-+)$ and $(+-)$, 
with complete (thick lined) and 
without spin correlations (thin lined).
In a) the effect of chargino polarization on the forward--backward asymmetry,
$A_{FB}=33\%$, is about a factor 10 and in b)
the effect of spin correlations on the
parallel--antiparallel asymmetry, $A_{PA}=-10\%$, is about a factor 2.
\la{fig_4}
}
\end{figure}

\begin{figure}[t]
\hspace{-.9cm}
\begin{minipage}{7cm}
\begin{picture}(7,5)
\put(-.2,0){\includegraphics{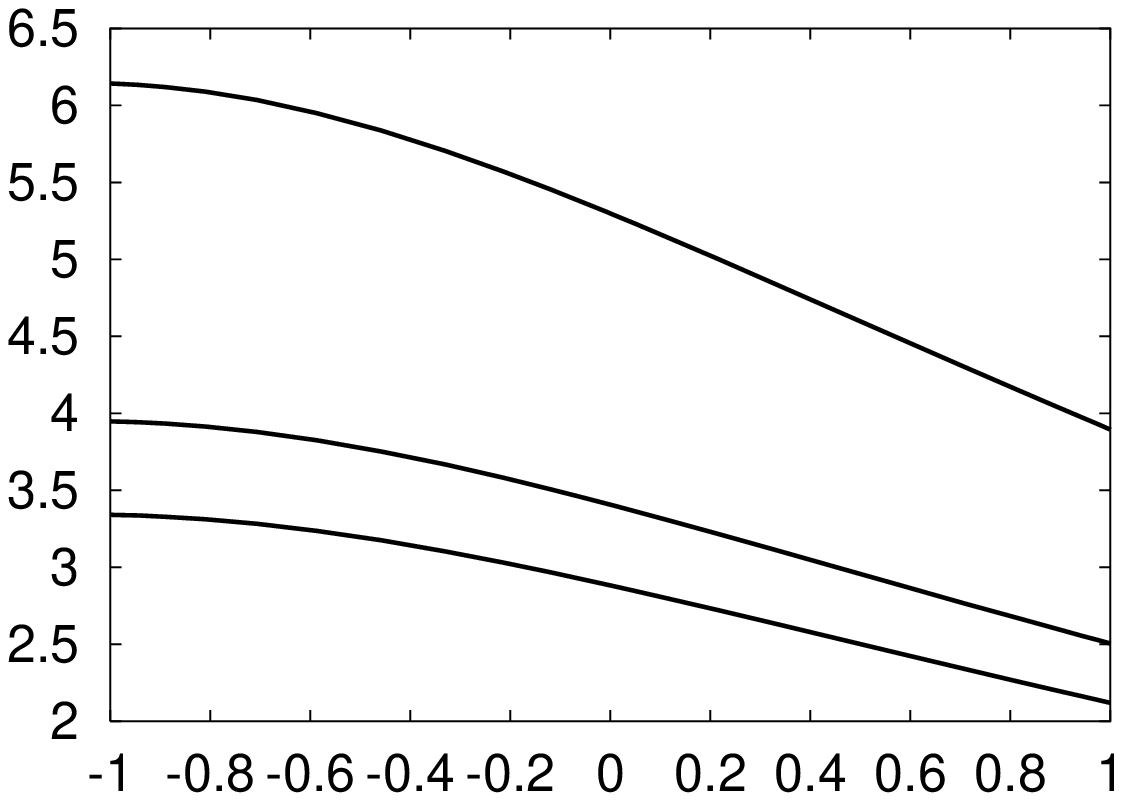}}
\put(4,4.2){\small $\sqrt{s}=230$~GeV}
\put(5.5,-.2){\small $\cos\Theta_{e^+ e^-}$}
\put(0,5.1){\small $d\sigma/d\cos\Theta_{e+e^-}$[fb]}
\put(1.3,2.7){\small $(+-)$}
\put(1.4,1.3){\small $(00)$}
\put(1.3,3.9){\small $(-+)$}
\put(3.5,5.2){\small a)}
\end{picture}\par\vspace{+.3cm}
\end{minipage}\hfill\hspace{.2cm}
\begin{minipage}{7cm}
\begin{picture}(7,5)
\put(-.2,0){\includegraphics{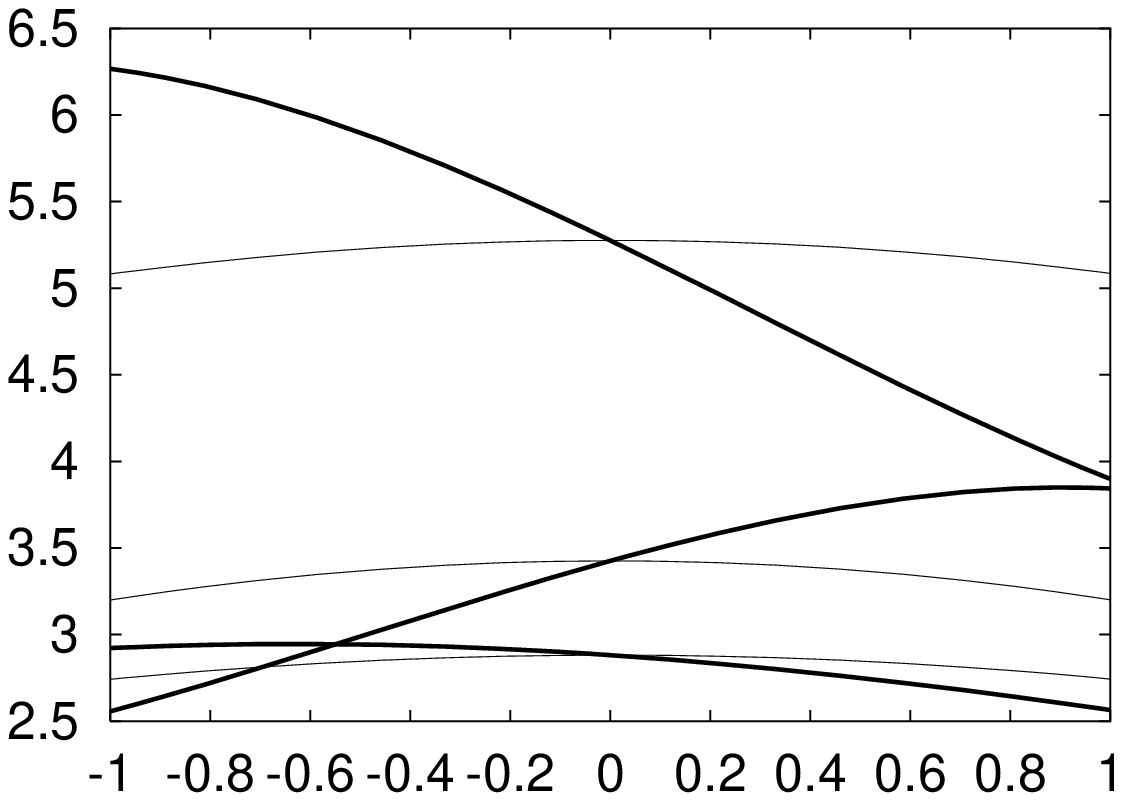}}
\put(4,4.2){\small $\sqrt{s}=230$~GeV}
\put(5.5,-.2){\small $\cos\Theta_{e^-}$}
\put(0,5.1){\small $d\sigma/d\cos\Theta_{e^-}$[fb]}
\put(5.5,1.7){\small $(+-)$}
\put(3.7,1.2){\small $(00)$}
\put(1.4,3.8){\small $(-+)$}
\put(3.5,5.2){\small b)}
\end{picture}\par\vspace{.3cm}
\end{minipage}
\caption{Neutralino production and decay,
$e^+ e^-\to \tilde{\chi}^0_1 \tilde{\chi}^0_2$, 
$\tilde{\chi}^0_2\to \tilde{\chi}^0_1 e^+ e^-$:
a) opening angle distribution 
of decay $e^+$ and $e^-$ and b) angular distribution
of decay $e^-$ at $\sqrt{s}=230$~GeV 
for unpolarized beams (00) and both 
beams polarized $P_{e^-}=\mp 85\%$, $P_{e^+}=\pm 60\%$, $(-+)$ and $(+-)$,
with complete (thick lined) and without (thin lined)
spin correlations between production and decay. 
In a) the distribution is exactly independent of spin correlations
due to the Majorana character. 
In b) the forward--backward asymmetry
reaches $A_{FB}=-12\%$ for $(-+)$, due to the spin correlations. 
\la{fig_5}
}
\vspace{9cm}
\end{figure}

\end{document}